\documentclass[aps,prb,amsmath,amssymb,reprint,longbibliography,superscriptaddress]{revtex4-2}
\usepackage[T2A]{fontenc}
\usepackage[utf8]{inputenc}
\usepackage[english]{babel}
\usepackage{mathtools}
\usepackage{graphicx}
\usepackage{mathrsfs}
\usepackage{accents}
\usepackage{amssymb}
\usepackage{xcolor}
\usepackage[shortlabels]{enumitem}
\usepackage[unicode=true,colorlinks=true,citecolor=blue,urlcolor=blue]{hyperref}
\usepackage{bm}
\usepackage{physics}
\usepackage[right=2cm, left=2cm, top=2cm]{geometry}
\renewcommand{\selectlanguage}[1]{}
\makeatletter\AtBeginDocument{\let\@elt\relax}\makeatother

\renewcommand{\i}{{i}}
\newcommand{\e}{\mathrm{e}}
\newcommand{\ttilde}[1]{\accentset{\approx}{#1}}

\begin{document}
\newcommand{\NL}[1]{\textcolor{blue}{#1}}
\newcommand{\commentNikita}[1]{\textcolor{blue}{\textbf{Nikita:} \textit{#1}}}
\newcommand{\addDima}[1]{\textcolor{red}{#1}}
\newcommand{\commentDima}[1]{\textcolor{red}{\textbf{Dima:} \textit{#1}}}

\title{Birefringent spin-photon interface generates polarization entanglement}
\author{Nikita Leppenen}
\email{nikita.leppenen@weizmann.ac.il}
\affiliation{Department of Chemical \& Biological Physics, Weizmann Institute of Science, Rehovot 7610001, Israel}

\author{Dmitry S. Smirnov}
\affiliation{Ioffe Institute, 194021 St. Petersburg, Russia}

\date{\today}
\begin{abstract}
  A spin-photon interface based on the luminescence of a singly charged quantum dot in a micropillar cavity allows for the creation of photonic entangled states. Current devices suffer from cavity birefringence, which limits the generation of spin-photon entanglement.
  % generation protocols require axially symmetric cavities, and the birefringence is considered a limitation.
  In this paper, we theoretically study the light absorption and emission by the interface with an anisotropic cavity and derive the maximal excitation and spin-photon entanglement conditions. We show that the concurrence of the spin-photon state equal to one and complete quantum dot population inversion can be reached for a micropillar cavity with any degree of birefringence by tuning the quantum dot resonance strictly between the cavity modes. This sweet spot is also valid for generating a multiphoton cluster state, as we demonstrate by calculating the three-tangle and fidelity with the maximally entangled state.
\end{abstract}

\maketitle
\twocolumngrid

\section{Introduction}

Spin entanglement is an essential resource for quantum information processing due to the excellent isolation of spin degree of freedom from the environment~\cite{dyakonov_book}. A basic example of the entanglement production is given by absorption of a linearly polarized photon in GaAs-based nanostructures when a Bell state of electron and hole spins is optically generated~\cite{ivchenko91,PhysRevLett.66.2491}. However, flying qubits naturally represented by photons are more useful for quantum information transfer. Their polarization entanglement is routinely produced by biexciton recombination~\cite{Akopian2006,Dousse2010,PhysRevLett.121.033902,PRXQuantum.3.020363}. 

Photons can be coupled to the localized qubits --- electron spins in quantum dots (QDs), by zero-dimensional spin-photon interfaces~\cite{uppu2021quantum}, which can be efficiently realized using micropillar cavities~\cite{nowak2014deterministic,PhysRevLett.116.020401,nowak2014deterministic,Tomm2021,RAKHLIN2023119496} (see Refs.~\onlinecite{senellart2017high,10.1063/PT.3.4962} for recent reviews). These interfaces allow for the photon polarization manipulation by electron spin~\cite{PhysRevB.78.085307,doi:10.1126/science.aat3581,mehdi2022controlling} and fast control of electron spin by photons~\cite{singleSpin,PhysRevLett.112.116802,PhysRevResearch.4.L042039,antoniadis2023cavity}. But most importantly, electron-photon entanglement can be created in these structures~\cite{DeGreve2012,Gao:2012fk,PhysRevLett.110.167401}.

An auspicious task is the generation of entangled photon cluster states~\cite{Briegel2009}. Among different approaches~\cite{PhysRevA.95.022304,istrati2020sequential,besse2020realizing,PhysRevA.105.L030601} including parametric down conversion~\cite{PhysRevLett.121.250505}, the most scalable one is the sequential entanglement of spin of a photon and localized electron~\cite{PhysRevLett.95.110503,lindner_proposal_2009,thomas2022efficient}. After the first realization of the theoretical proposal~\cite{Schwartz434}, the intensive investigations in this direction continue until now~\cite{cogan_deterministic_2023-1,coste_high-rate_2023}.

%Quantum dots constitute a fundamental component in photonic quantum information technologies~\cite{heindel_quantum_2023}. Experimentally realized spin-photon interface based on the fluorescence of the single charged quantum dot~\cite{cogan_deterministic_2023-1,coste_high-rate_2023} (QD), allowing to create entangled states of photons such as GHZ states or cluster states. About single photon protocol~\cite{hermans_entangling_2023}

An important factor limiting the generation of many entangled photons is the intrinsic electron spin relaxation~\cite{book_Glazov}. An approach for its investigation by measurement of the equilibrium electron spin noise through the photon correlation functions was proposed theoretically~\cite{BackAction,leppenen_quantum_2021} and recently realized experimentally~\cite{gundin2024spin}. However, a significant obstacle for the spin dynamics investigation and spin-photon entanglement generation remains due to the splitting or birefringence of the micropillar cavity modes~\cite{PhysRevLett.102.097403,kim2013quantum,tiecke2014nanophotonic,Waks2016,10.1063/1.5026799,PhysRevLett.116.020401,mehdi2022controlling,antoniadis2023cavity} despite multiple attempts to reduce it~\cite{PhysRevLett.102.097403,Arnold2015,doi:10.1021/acsphotonics.8b01380,PhysRevApplied.11.061001,Galimov2021}.

\begin{figure}[t!]
  \centering
  \includegraphics[width=\linewidth]{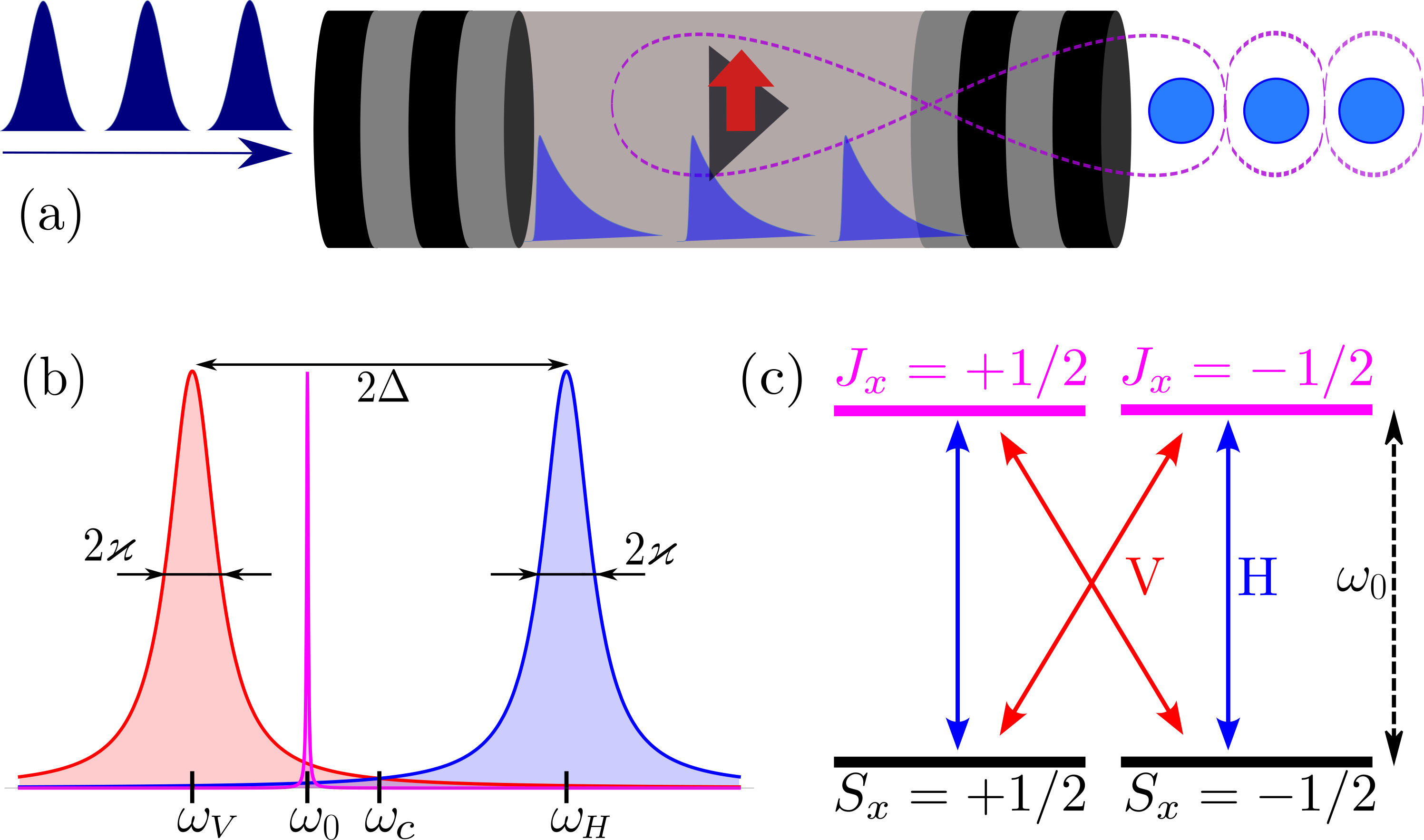}
  \caption{(a) Sketch of the system: quantum dot (black pyramid) with an electron with a spin (red arrow) in a micropillar cavity. Short optical pump pulses (dark blue) transform inside the cavity into exponentially decaying ones (lilac) and excite trion in the quantum dot. The photons emitted after the trion recombination (cyan) are polarization entangled with each other and the electron spin. (b) The spectrum of $H$ (blue) and $V$ (red) cavity modes and the quantum dot resonance (magenta). (c) Energy levels of the quantum dot (black and magenta lines) and transitions between them involving $H$ (blue arrows) and $V$ (red arrows) photons.}
  \label{fig:model}
\end{figure}

In this work, we theoretically address this issue. After formulating the model in Sec.~\ref{sec:model}, we show proof of principle numerical calculation in Sec.~\ref{sec:num} and then describe optical excitation of the quantum dot in Sec.~\ref{sec:exc}, spin-photon entanglement generation in Sec.~\ref{sec:rec}, and multiphoton entanglement in Sec.~\ref{sec:many}. In conclusion (Sec.~\ref{sec:conclusion}), we stress that there is a ``sweet spot'' at which trion can be efficiently excited and complete spin-photon entanglement can be realized for arbitrary strength of the micropillar cavity birefringence.

\section{Model}
\label{sec:model}

We describe a micropillar cavity with an electron-charged quantum dot [Fig.~\ref{fig:model}(a)] by the following general Hamiltonian~\cite{singleSpin,BackAction,leppenen_quantum_2021}:
\begin{equation}
  \label{eq:ham}
  {\cal H} = \sum_{H,V} {\cal H}^{\text{(cav)}}_{H,V}+\sum_{\pm}\qty[{\cal H}^{\text{(trion)}}_\pm+{\cal H}^{\text{(int)}}_\pm+{\cal H}^{\text{(exc)}}_\pm].
\end{equation}
Let us describe each of these terms. The first one reads ${\cal H}^{\text{(cav)}}_{H,V} = \omega_{H,V} c^\dagger_{H,V}c_{H,V}$. It describes the energy of photons in two linearly polarized cavity modes, H and V, with the corresponding annihilation operators $c_{H,V}$ and frequencies $\omega_{H,V}$ (hereafter, we set $\hbar=1$). It is convenient to represent these frequencies as $\omega_{H,V}=\omega_c\pm\Delta$, where $\omega_c$ is the central frequency and $2\Delta$ is the splitting between cavity modes, as shown in Fig.~\ref{fig:model}(b).

Absorption of a photon by a quantum dot with an electron leads to the formation of a negatively charged trion. Its energy is given by ${\cal H}^{\text{(trion)}}_\pm = \omega_0 a^\dagger_{\pm 3/2} a_{\pm 3/2}$, where $\omega_0$ is the trion resonance frequency and $a_{\pm 3/2}$ are the annihilation operators of trions with the heavy hole spin $\pm3/2$, which correspond to the trion pseudospin $J_z=\pm1/2$.

The third term in Eq.~\eqref{eq:ham} 
\begin{equation}
    {\cal H}^{\text{(int)}}_\pm = g\qty[a^\dagger_{\pm 3/2} c_\pm a_{\pm 1/2}+a^\dagger_{\pm 1/2} c_\pm^\dagger a_{\pm 3/2}],
\end{equation}
describes light-matter coupling with the strength $g$ in a circular basis with
\begin{equation}
  \label{eq:c_x_y}
  c_\pm=\frac{c_H\pm\i c_V}{\sqrt{2}}.
\end{equation}
The operators $a_{\pm1/2}$ destroy an electron in the ground state with the spin $S_z=\pm1/2$. The selection rules described by this Hamiltonian can also be rewritten in the basis of H and V modes for the electron and trion states $S_x,J_x=\pm1/2$. The corresponding transitions are shown in Fig.~\ref{fig:model}(c).

The last term of the general Hamiltonian ${\cal H}^{\text{(exc)}}_\pm = {\cal E}_\pm^*(t)c_\pm+{\cal E}_\pm(t)c^\dagger_\pm$ stands for the resonant cavity excitation by external light pulse described by the amplitudes ${\cal E}_\pm(t)$ of two circular components. A train of pulses is needed to generate many entangled photons. But if the intervals between the pulses are long enough (longer than the trion recombination time $\gamma^{-1}$), one can separately consider each pulse. We assume the length of the pulses to be so short that their spectrum covers both cavity modes, and take ${\cal E}_\pm(t)={\cal E}_\pm\delta(t)$ in what follows. Note that in the opposite limit of the pump pulses longer than the inverse cavity decay rate, the birefringence does not play a significant role; it only leads to the polarization-dependent photon injection rate.

We consider only one, the most important incoherent process, the escape of photons from the cavity with the polarization-independent rate $2\varkappa$. It is assumed to be larger than the coupling strength, $\varkappa\gg g$. This is the so-called fast cavity regime, which is realized in most structures. Typically, in the experiments, the decay rates of the two cavity modes are similar, so we leave the study of the effect of their difference for future works. We neglect the incoherent trion decay, assuming it to be slower than $\gamma$. For this reason, the trion state is represented by a narrow Lorentzian in Fig.~\ref{fig:model}(b). The transmission spectrum of the system calculated within the input-output approach is shown in Appendix~\ref{sec:Trans}. In experiments, the reflection geometry is commonly employed. Moving forward, our focus is on studying spin entanglement with the photons emitted from the microcavity, with the emission direction being inconsequential for the theoretical analysis.

The dynamics of the system can be described on equal footing by the Lindblad equation or by the method of quantum trajectories~\cite{molmer_monte_1993,carmichael_open_2009}. In the latter approach, the non-Hermitian Hamiltonian should be considered
\begin{equation}
  \label{eq:H_nh}
    {\cal H}_{\rm nh} = {\cal H}-i\varkappa \sum_{\pm} c_\pm^\dagger c_{\pm}
\end{equation}
and the jump operators are $C_\pm = \sqrt{2\varkappa} c_\pm$. Averaging over individual quantum trajectories gives the expectation values of physical observables in the system.

\section{Numerical calculation}
\label{sec:num}

\begin{figure}[t!]
  \centering
  \includegraphics[scale = 0.57]{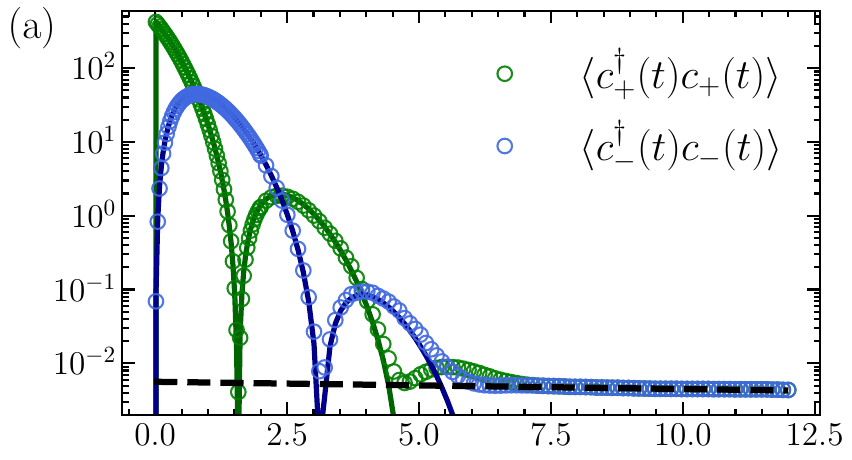}
  \includegraphics[scale = 0.57]{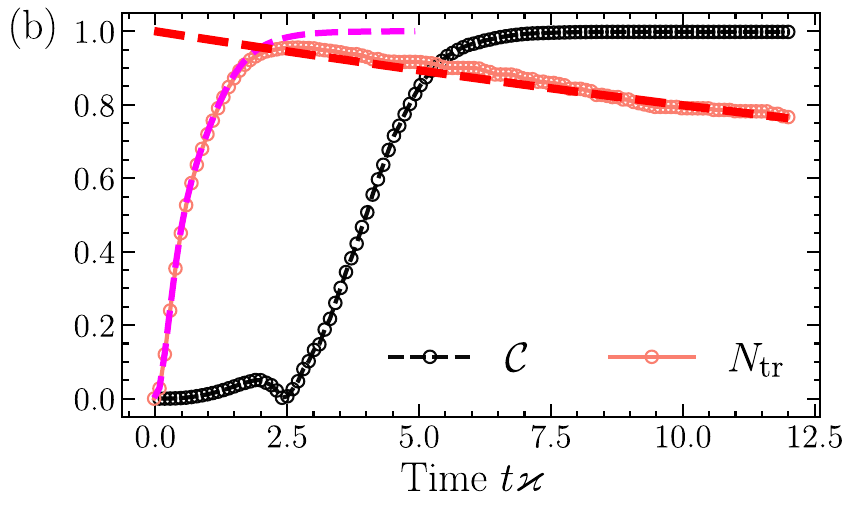}
  \caption{(a) Numbers of $\sigma^\pm$ photons in the cavity mode (open symbols). Solid green and blue lines are calculated after Eqs.~\eqref{eq:c_x_y} and~\eqref{eq:ct}. The black dashed line is calculated after Eq.~\eqref{eq:photons}. (b) The population of the trion state and electron-photon concurrence as functions of time. Magenta and red dashed lines are calculated after Eqs.~\eqref{eq:tr_sol} and~\eqref{eq:gamma}, respectively. The parameters of the calculation are $g=0.15\varkappa$, $\Delta=\varkappa$, $\omega_0=\omega_c$, ${\cal E}_-=0$, ${\cal E}_+=\pi\varkappa/g$ (a finite pulse duration of $0.003/\varkappa$ is used).}
  \label{fig:num_res}
\end{figure}

To illustrate the following analytical theory, we present a numerical calculation for a particular set of parameters. In Fig.~\ref{fig:num_res}, we consider a $\sigma^+$ circularly polarized excitation and initial electron spin polarization along the $x$ axis, $S_x=1/2$. The other parameters are given in the figure caption. For the simulations, we use the quantum Monte-Carlo trajectory method~\cite{molmer_monte_1993} realized in QuTiP python framework~\cite{johansson_qutip_2013} and perform averaging over 300 individual quantum trajectories.

Panel (a) shows the numbers of $\sigma^+$ and $\sigma^-$ polarized photons in the cavity mode as functions of time. The beatings with the period $\pi/\Delta$ are related to the splitting between linearly polarized cavity modes. After the initial injection of photons, their number exponentially decays due to the escape of photons from the cavity.

During this time, a trion gets excited in the quantum dot. Its population $N_{\rm tr}=\expval{a_{+3/2}^\dag a_{+3/2}+a_{-3/2}^\dag a_{-3/2}}$ is shown in Fig.~\ref{fig:num_res}(b) by the red circles. It starts from zero and then quickly ($t \sim 1/\varkappa$) increases almost to one, corresponding to the $\pi$ pulse. Then, the trion starts to decay slowly, giving rise to a small population of the cavity mode at the long time scales. The black circles show the concurrence between a photon emitted from the cavity at the given time and the spin of the electron left in the quantum dot. It is close to one, which demonstrates complete spin-photon entanglement.

The analytical expressions for these quantities are given below. However, the main results can already be seen from the figure: (i) Despite splitting of the cavity modes and their detuning from the trion resonance, a complete population inversion ($\pi$ pulse) is possible. (ii) Even strong birefringence of the microcavity does not spoil spin photon entanglement so that concurrence reaches one. Below, we describe both results analytically and in more detail.

\section{Trion excitation}
\label{sec:exc}

Light quantization does not play a role in the weak coupling regime ($g\ll\varkappa$). Then, the average amplitudes of the linearly polarized light components in the cavity obey the classical equations
\begin{equation}
  \label{eq:ct}
  \expval{c_{H,V}(t)} = \expval{c_{H,V}(0)}e^{-i\omega_{H,V}t-\varkappa t},
\end{equation}
with the initial conditions determined by the pump pulse: $\expval{c_{H}(0)} = \frac{i}{\sqrt{2}}({\cal E}_++{\cal E}_-)$, $\expval{c_V(0)} = \frac{1}{\sqrt{2}}({\cal E}_+-{\cal E}_-)$. These expressions are used to calculate the blue and green solid lines in Fig.~\ref{fig:num_res}(a). The total number of photons decays at the rate of $2\varkappa$.

The trion decay can be neglected during the photon lifetime in the cavity so that the trion excitation can be described by the Schrodinger equation~\cite{yugova_pump-probe_2009}. The quantum dot state is characterized by the four quantum amplitudes $\psi_{\pm1/2}(t)$ and $\psi_{\pm3/2}(t)$ of the corresponding electron and trion spin states. From the Hamiltonian~\eqref{eq:ham}, one can see that the Schrodinger equation for them reads
\begin{subequations}
  \label{eq:trion}
  \begin{equation}
    \i\dot{\psi}_{\pm 3/2}=\omega_0\psi_{\pm3/2}+g\expval{c_\pm(t)}\psi_{\pm1/2},
  \end{equation}
  \begin{equation}
    \i\dot{\psi}_{\pm 1/2}=g\expval{c_\pm^\dag(t)}\psi_{\pm3/2},
  \end{equation}
\end{subequations}
where the dot denotes the time derivative.  One can see that the excitations of the trions with $J_z=\pm1/2$ are independent of each other. The solution of these equations for $t\to\infty$ gives the trion state as a function of the initial electron state, amplitude, and polarization of the exciting light.

In the general case, an analytical solution of these equations is complicated. We solve them numerically for the different splittings of the cavity modes, $\Delta$, and different trion resonance frequencies, $\omega_0$. For each $\Delta$ and $\omega_0$, we find the maximum (as a function of the pump amplitudes ${\cal E}_+$ and ${\cal E}_-$) occupancy of the trion state $N_{\rm tr}^{\rm max}$ for the initial electron in-plane state. This gives the average probability of excitation of the two trion states from the corresponding electron states.

% \begin{figure}
%   \centering
%   \includegraphics[width=0.9\linewidth]{img/Excitation2.pdf}
%   \caption{Maximal possible trion population as a function of the trion resonance frequency and splitting between cavity modes for the initial in-plane electron spin orientation. The cyan lines show the levels of $N_{\rm tr}^{\rm max}=0.9$. \commentDima{Improve labels, fix italics.}}
%   \label{fig:trion_exc}
% \end{figure}

\begin{figure*}[t!]
        \centering
        \includegraphics[scale = 0.8]{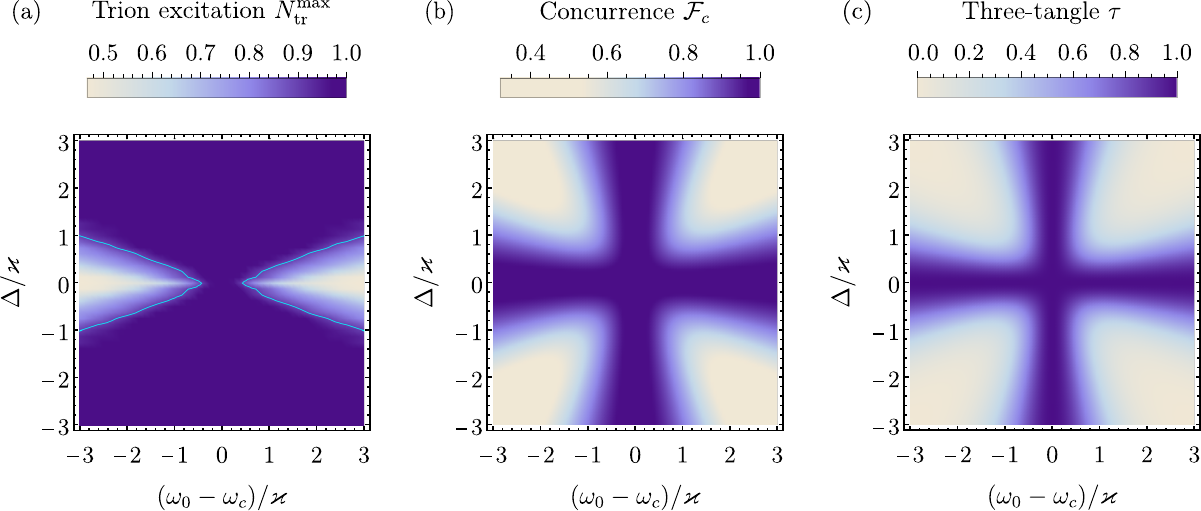}
        \caption{(a) Maximal possible trion population as a function of the trion resonance frequency and splitting between cavity modes for the initial in-plane electron spin orientation obtained from the numerical solution of Eqs.~\eqref{eq:trion}. The cyan lines show the levels of $N_{\rm tr}^{\rm max}=0.9$. (b) Electron-photon concurrence calculated after Eq.~\eqref{eq:F_c} (c) Three-tangle of the electron and two photons in the cluster-like state calculated after Eq.~\eqref{eq:tau}.  }
        \label{fig:maps} 
\end{figure*}

The result is shown in Fig.~\ref{fig:maps}(a). Surprisingly, the trion can be excited with the probability close to one for the most range of the parameters $\Delta$ and $\omega_0$. The cyan curves show the levels of $N_{\rm tr}^{\rm max}=0.9$. Several limiting cases can explain this result.

First, if the splitting between the modes is zero, then the quantum dot represents a pair of two-level systems. If the quantum dot frequency coincides with the cavity frequency, $\omega_0=\omega_c$, the populations of the trion states experience Rabi oscillations as functions of the pump amplitudes~\cite{yugova_pump-probe_2009}:
\begin{equation}
  \expval{a_{\pm3/2}^\dag a_{\pm 3/2}}=\sin^2\left(g|{\cal E}_\pm|/\varkappa\right).
\end{equation}
So the $\pi$ pulse is realized, for example, by a linearly polarized excitation with ${\cal E}_\pm=\pi\varkappa/(2g)$. If the trion resonance is detuned from the cavity frequency, one can show that the maximum trion population is given by
\begin{equation}
  N_{\rm tr}^{\rm max}=\frac{1}{2}\left[1+\sech\left(\frac{\pi(\omega_0-\omega_c)}{2\varkappa}\right)\right].
\end{equation}
This expression describes the trion population along the line $\Delta=0$ in Fig.~\ref{fig:maps}(a). For small trion detunings, it reaches one, and for large detunings, it tends to one-half. The trion population for the strong pump pulses experiences the Rabi oscillations from zero to one. Still, when the light amplitude inside the cavity decays according to Eq.~\eqref{eq:ct}, the amplitude of the oscillations decreases, so a finite trion population $N_{\rm tr}<1$ is left in the QD after the escape of all photons from the cavity.

Another simple limit is realized for $\omega-\omega_c=\pm\Delta$ [diagonals of Fig.~\ref{fig:maps}(a)]. The trion resonance coincides, in this case, with one of the two cavity modes. So, the pump polarization along this mode can realize the complete population inversion. Now, it becomes natural that in the whole region of $|\omega-\omega_c|<|\Delta|$, the $\pi$ pulses can be realized for some amplitudes and polarizations of the pump pulse.

The most important case is the trion resonance exactly between the cavity modes, $\omega_0=\omega_c$. Due to the detuning from both modes, one could expect only a limited maximum possible trion population. However, we find that if the pump amplitudes ${\cal E}_+$ and ${\cal E}_-$ are shifted in phase by $\pi/2$ (${\cal E}_+=|{\cal E}_+|$, ${\cal E}_-=-\i|{\cal E}_-|$), the solution of Eq.~\eqref{eq:trion} reads
\begin{subequations}
  \label{eq:tr_sol}
   \begin{equation}
     \psi_{\pm 3/2}(t) = \psi_{\pm 1/2}(0)\sin(g \tilde{\cal E}_\pm(t)\over \Delta^2+\varkappa^2),
   \end{equation}
   where
\begin{multline}
   \tilde{\cal E}_\pm(t) = \abs{{\cal E}_\pm}\varkappa\mp\abs{{\cal E}_{\mp}}\Delta+\left[(\pm\abs{{\cal E}_\mp}\Delta-\abs{{\cal E}_\pm}\varkappa)\cos(\Delta t)\right.  \\ \left.  +(\abs{{\cal E}_\pm}\Delta+\abs{{\cal E}_\mp}\varkappa)\sin(\Delta t)\right]e^{-\varkappa t}.
   % \tilde{\cal E}_\pm = (\abs{{\cal E}_\pm}\varkappa\mp\abs{{\cal E}_{\mp}}\Delta)+e^{-\varkappa t}\left[(\pm\abs{{\cal E}_\mp}\Delta-\abs{{\cal E}_\pm}\varkappa)\right.\times  \\ \left.\times \cos(\Delta t) +(\abs{{\cal E}_\pm}\Delta+\abs{{\cal E}_\mp}\varkappa)\sin(\Delta t)\right].
\end{multline}
\end{subequations}
From this, one can see the conditions for complete trion excitation ($\pi$ pulses):
\begin{equation}
  \label{eq:pi_pulse}
 \sin^2\qty[g\frac{\abs{{\cal E}_\pm}\varkappa\mp\abs{{\cal E}_\mp}\Delta}{\Delta^2+\varkappa^2}]=  1.
\end{equation}
This proves that $N_{\rm tr}^{\rm max}=1$ along the vertical line in the middle of Fig.~\ref{fig:maps}(a), and this will be important for the spin-photon entanglement in the next section.

For example, at $\Delta=\varkappa$, Eq.~\eqref{eq:pi_pulse} is fulfilled for circularly polarized light with ${\cal E}_-=0$, ${\cal E}_+=\pi\varkappa/g$. This particular case was used for the numerical calculations in Fig.~\ref{fig:num_res}.

Finally, we note that the $\pi$ pulses completely transfer populations of electron spin-up and spin-down states to the trion spin-up and -down states, respectively. This means that after excitation, the trion spin $J_z$ equals the initial electron spin $S_z$. Also, for the electron in-plane spin orientation, the trion spin will also be completely polarized in the $(xy)$ plane.

\section{Spin-photon entanglement}
\label{sec:rec}

Having described the trion excitation at the short time scales, we now analyze the slow trion radiative recombination and the spin-photon entanglement generation. Making use again of the weak coupling regime, we use the Schr\"odinger equation with the non-Hermitian Hamiltonian~\eqref{eq:H_nh} to describe the trion recombination~\cite{PhysRevB.78.085307,BackAction}. We consider a trion state described by the amplitudes $\psi_{+3/2,0}$ and $\psi_{-3/2,0}$ (the subscript $0$ indicates the absence of photons in the cavity mode). Then the Schr\"odinger equation gives us
\begin{subequations}
  \begin{equation}
    \i\dot{\psi}_{\pm1/2,\pm}=\left(\omega_c-\i\varkappa\right)\psi_{\pm1/2,\pm}+\Delta\psi_{\pm1/2,\mp}+g\psi_{\pm3/2,0},
  \end{equation}
  \begin{equation}
    \i\dot{\psi}_{\pm1/2,\mp}=\left(\omega_c-\i\varkappa\right)\psi_{\pm1/2,\mp}+\Delta\psi_{\pm1/2,\pm},
  \end{equation}
\end{subequations}
where the first subscript $\pm1/2$ in the wave functions refers to the electron spin, and the second subscript $\pm$ denotes a single $\sigma^\pm$ photon in the cavity.

The solution of these equations reads
\begin{subequations}
  \label{eq:wf}
  \begin{equation}
    \psi_{\pm1/2,\pm}=\frac{g(\omega_0-\omega_c+\i\varkappa)\psi_{\pm3/2,0}}{(\omega_0-\omega_c+\i\varkappa)^2-\Delta^2},
  \end{equation}
  \begin{equation}
    \psi_{\pm1/2,\mp}=\frac{g\Delta\psi_{\pm3/2,0}}{(\omega_0-\omega_c+\i\varkappa)^2-\Delta^2},
  \end{equation}
\end{subequations}
where we took into account that the trion wave functions $\psi_{\pm3/2,0}$ have a phase factor of $\e^{-\i\omega_0 t}$. In Appendix~\ref{sec:Lindblad}, we show that the same result can be obtained from the solution of the Lindblad equation for the density matrix of the system.

The numbers of photons in the cavity modes are given by 
\begin{equation}
  \label{eq:photons}
  \expval{c_\pm^\dag c_\pm}=|\psi_{+1/2,\pm}^2|+|\psi_{-1/2,\pm}^2|,
\end{equation}
and this expression is shown by the black dashed line in Fig.~\ref{fig:num_res}(a). It allows us to calculate the trion (amplitude) decay rate as:
\begin{equation}
  \gamma=\varkappa\frac{|\psi_{\pm1/2,+}^2|+|\psi_{\pm1/2,-}^2|}{|\psi_{\pm 3/2,0}^2|},
\end{equation}
because it equals the photon escape rate. Thus we obtain
\begin{equation}
  \label{eq:gamma}
  \gamma=\frac{1}{2}\sum_{H,V}\frac{g^2\varkappa}{(\omega_0-\omega_{H,V})^2+\varkappa^2},
\end{equation}
which is generally of the order of $g^2/\varkappa$ or smaller. The slow trion decay $N_{\rm tr}=\exp(-2\gamma t)$ is shown by the red dashed line in Fig.~\ref{fig:num_res}(b).

Because of the assumption of equal decay rates of the two cavity photon modes, the trion decay rate does not depend on the trion spin. So, the wave function of an electron in the quantum dot and an emitted photon is also given by Eqs.~\eqref{eq:wf}. It can be written (up to a phase factor) as
\begin{equation}
  \label{eq:e-ph}
  \Psi=\psi_{+3/2,0}\ket{\uparrow,\tilde{+}}+\psi_{-3/2,0}\ket{\downarrow,\tilde{-}},
\end{equation}
where $\uparrow$ and $\downarrow$ denote the electron spin state, and the photon states are
\begin{equation}
  \label{eq:tilde_pm}
  \ket{\tilde{\pm}}=\cos\alpha\ket{\pm}-\i\sin\alpha\e^{\i\beta}\ket{\mp}
\end{equation}
with $\ket{\pm}$ denoting $\sigma^\pm$ photons and
\begin{subequations}
  \begin{equation}
    \tan\alpha=\frac{\Delta}{\sqrt{(\omega_0-\omega_c)^2+\varkappa^2}},
  \end{equation}
  \begin{equation}
    \tan\beta=\frac{\omega_0-\omega_c}{\varkappa}.
  \end{equation}
\end{subequations}
Both angles $\alpha$ and $\beta$ are in the range $\left(-\pi/2,\pi/2\right)$ and depend on the trion resonance frequency and cavity mode splitting.

% \begin{figure}[htp]
%     \centering
%     \includegraphics[scale = 0.45]{img/conc_plot_1a.png}
%     \includegraphics[scale = 0.45]{img/three_tangle_plot.png}
%     \caption{\commentDima{Fix fonts, add labels, make caption.}}
%     \label{fig:concurrence}
% \end{figure}

We stress that the electron-photon state is pure. Moreover, it has the same form as for the optically isotropic cavity~\cite{lindner_proposal_2009} (with $\Delta=0$) but with the rotated photon states~\eqref{eq:tilde_pm}. So, we can easily calculate spin-photon concurrence
  \begin{equation}
    \mathcal C=2\sqrt{J_x^2+J_y^2}\mathcal F_c
  \end{equation}
with
  \begin{equation}
  	\label{eq:F_c}
    \mathcal F_c=\sqrt{1-\sin^2(2\alpha)\sin^2\beta}.
  \end{equation}
  One can see that the spin-photon entanglement requires in-plane trion spin polarization as usual~\cite{lindner_proposal_2009}. The cavity birefringence is accounted for by the factor $\mathcal F_c\le1$, shown in Fig.~\ref{fig:maps}(b) as a function of the system parameters. It is related to the scalar product of the two modified photon states as $\mathcal F_c^2=1-|\left<\tilde{+}\middle|\tilde{-}\right>^2|$, so the largest concurrence requires the minor overlap of the emitted photon states. The overlap vanishes in the absence of cavity mode splitting, $\Delta=0$, and one arrives at $\mathcal C=2\sqrt{J_x^2+J_y^2}$~\cite{leppenen_quantum_2021}.

  But strikingly, at $\omega_0=\omega_c$ we obtain $\beta=0$, $\left<\tilde{+}\middle|\tilde{-}\right>=0$, and $\mathcal F_c=1$ also. So, despite the cavity birefringence, the perfect spin-photon entanglement is achieved when the trion resonance is located exactly between the two cavity modes. Qualitatively, perfect spin-photon entanglement requires zero overlap between two possible emitted photon states $\ket{\tilde{\pm}}$, which is possible for the equal probabilities of photon emission to H and V modes only. From Eq.~\eqref{eq:gamma}, one can see that this corresponds exactly to $\omega_0=\omega_c$, which was used already in Ref.~\onlinecite{coste_high-rate_2023}.

We note that complete trion excitation and in-plane spin polarization are also realized in this case, as shown in the previous section. For this reason, we call the condition $\omega_0=\omega_c$ a ``sweet spot'' for the entanglement generation in the birefringent microcavities. This is the main result of our work.
  
\section{Multiphoton entanglement}
\label{sec:many}

Spin-photon entanglement is useful by itself and indispensable for generating many-body entangled photon cluster states. The form of the electron-photon wave function~\eqref{eq:e-ph} shows that the protocol of Ref.~\onlinecite{lindner_proposal_2009} can be successfully applied to the birefringent micropillar cavities. But in the result, the circularly polarized $\sigma^\pm$ photon states should be replaced with $\ket{\tilde{\pm}}$.

For example, an electron spin rotation by $\pi/2$ between the two excitation pulses leads to the entangled state of an electron and two photons
\begin{equation}
  \Psi_3=\frac{\ket{\uparrow\tilde{+}\tilde{+}}+\ket{\downarrow\tilde{-}\tilde{+}}+\ket{\downarrow\tilde{-}\tilde{-}}-\ket{\uparrow\tilde{+}\tilde{-}}}{2}.
\end{equation}
The straightforward calculation shows that the three-tangle~\cite{PhysRevA.61.052306} of this state is given by
\begin{equation}
	\label{eq:tau}
  \tau=\mathcal F_c^4.
\end{equation}
It is shown in Fig.~\ref{fig:maps}(c). It has more narrow regions with high entanglement than the concurrence in panel (b) but generally has the same form. In particular, it reaches one at the sweet spot $\omega_0=\omega_c$.

The three tangle does not reach zero for any system parameters. This suggests that the cluster-like states generated by the birefringent cavity are always genuine entangled. In particular, localizable entanglement~\cite{PhysRevLett.92.027901} between any two emitted photons in the cluster state equals $\mathcal F_c$. In this sense, the entanglement length of the multiphoton state is infinite for any system parameters.

To quantify the degree of entanglement, we introduce the orthogonal photon states
\begin{equation}\label{eq:ttildepm}
  \ket{\ttilde{\pm}}=\cos(\theta/2)\ket{\pm}-\i\sin(\theta/2)\ket{\mp},
\end{equation}
where the angle $\theta$ is defined by
\begin{equation}
  \tan\theta=\frac{2\varkappa\Delta}{(\omega_0-\omega_c)^2+\varkappa^2-\Delta^2}.
\end{equation}
These states are the closest orthogonal states to $\ket{\tilde\pm}$ (see Appendix~\ref{sec:Stokes}), so they determine the optimal basis for the measurement of multiphoton entanglement.

We denote the cluster-like states with $n$ photons produced by the birefringent cavity as $\Psi_n$, and the ideal cluster states in the basis of $\ket{\ttilde{\pm}}$ states as $\Psi_n^{(0)}$. Then, the fidelity between these states is
\begin{equation}
  \label{eq:Fn}
  \left|\left<\Psi_n\middle|\Psi_n^{(0)}\right>\right|^2=\left(\frac{1+\mathcal F_c}{2}\right)^n.
\end{equation}
In particular, for $n=1$, the fidelity equals $(1+\mathcal F_c)/2$ in agreement with the general relations between concurrence and fidelity~\cite{PhysRevA.66.022307,PhysRevA.70.032326}. Eq.~\eqref{eq:Fn} shows that the fidelity exponentially decreases with the number of photons, except for the sweet spot, where the maximally entangled multiphoton states are generated.

\section{Conclusion}
\label{sec:conclusion}

We have developed a microscopic theory of the optical resonant excitation of a charged quantum dot in a birefringent micropillar cavity and its radiative recombination. The results show that a sweet spot exists for the system when the frequency of the trion resonance is placed exactly between the two cavity modes. At this point, even for the large cavity modes splitting, a complete trion excitation ($\pi$ pulse) and in-plane spin polarization are possible, as well as the complete spin-photon entanglement after trion recombination. It shows the way to increase the fidelity of the present spin-photon interfaces.

\begin{acknowledgements}
We thank A. A. Toropov for fruitful discussions and the Foundation for the Advancement of Theoretical Physics and Mathematics ``BASIS.'' The derivation of the analytical expressions for the spin-photon entanglement by D.S.S was supported by the Russian Science Foundation Grant No. 21-72-10035.
\end{acknowledgements}

\appendix

\section{Transmission spectra}
\label{sec:Trans}
Here, we study the transmission of the continuous wave coherent light with frequency $\omega$ through the birefringent spin-photon interface. We modify the input-output relations~\cite{walls_quantum_2008} for the case of the asymmetric cavity using the Hamiltonian~\eqref{eq:ham}
\begin{equation}
\label{eq:in_out}
    \begin{cases}
        \dot{\hat{c}}_+ = -i\omega_c \hat{c}_+-i\Delta \hat{c}_--\varkappa \hat{c}_+-ig \hat{a}_{+1/2}^\dagger \hat{a}_{3/2}+\sqrt{\varkappa} \hat{c}^{\rm in}_+\\
        \dot{\hat{c}}_- = -i\omega_c \hat{c}_--i\Delta \hat{c}_+-\varkappa \hat{c}_-+\sqrt{\varkappa}\hat{c}^{\rm in}_-\\
        \dot{\hat{a}}_{+3/2} = -i\omega_0 \hat{a}_{+3/2}-ig \hat{c}_+ \hat{a}_{+1/2}\\
        \dot{\hat{a}}_{+1/2} = -ig \hat{c}^\dagger_+\hat{a}_{+3/2}\\
        \hat{c}_{+}^{\rm out} = \sqrt{\varkappa} \hat{c}_+\\
        \hat{c}_{-}^{\rm out} = \sqrt{\varkappa} \hat{c}_-
    \end{cases}.
\end{equation}
Here $\hat{c}^{\rm in,out}_{\pm}$ are input and output $\sigma^{\pm}$ light field operators, which are proportional to $\e^{-\i\omega t}$, and we assume the electron and trion spin-down states to be not populated. These equations can be solved in the first order in the amplitude of the incident light by replacing $a_{+1/2}^\dag a_{+1/2}$ and $a_{+3/2}^\dag a_{+1/2}$ with one and zero, respectively~\cite{PhysRevB.78.085307}.

Without the cavity mode splitting, $\Delta = 0$, we arrive at the previously established results~\cite{PhysRevB.78.085307,singleSpin,leppenen_quantum_2021}: Transmission of $\sigma^-$ polarized light is given by
\begin{equation}
	t_0 = \frac{\hat{c}_-^{\rm out}}{\hat{c}_-^{\rm in}} = \frac{i \varkappa}{\omega-\omega_c+i\varkappa},
\end{equation}
which corresponds to the transmission when the optical selection rules forbid photon interaction with the QD, while for the $\sigma^+$ light polarization, we have 
\begin{equation}
	t_1 = \frac{\hat{c}_+^{\rm out}}{\hat{c}_+^{\rm in}}  = \frac{i\varkappa}{\omega-\omega_c+i\varkappa-\frac{g^2}{\omega-\omega_0}},
\end{equation}
relevant for the case when the QD can absorb the circularly polarized photons.

In the birefringent cavite case, $\Delta \neq 0$, from the solution of Eqs.~\eqref{eq:in_out} we obtain 
\begin{multline}
\label{eq:trans_full}
	t_{++} = \frac{t_1}{1+\Delta^2t_1t_0/\varkappa^2},\quad t_{--} = \frac{t_0}{1+\Delta^2t_1t_0/\varkappa^2}, \\ t_{-+} = -\frac{\i\Delta}{\varkappa}t_{++}t_0,\qquad t_{+-} = -\frac{i\Delta}{\varkappa}t_{--}t_1,
\end{multline}
where $t_{\alpha \beta} = \hat{c}_\beta^{\rm out}/\hat{c}_\alpha^{\rm in}$ for $\alpha ,\beta = \pm$. As one can see, the mode splitting results in the light polarization conversion. We note that these expressions are valid in both weak and strong coupling regimes.

\begin{figure}[t!]
  \centering
  \includegraphics[width=0.9\linewidth]{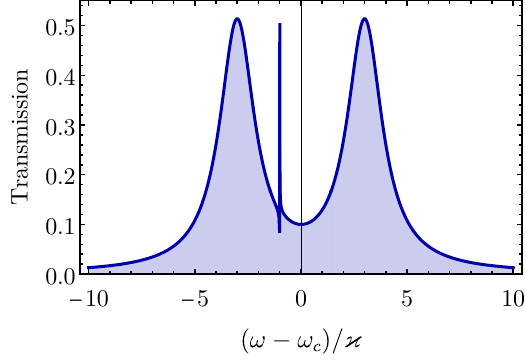}
  \caption{Intensity transmission spectrum for the unpolarized light calculated after Eq.~\eqref{eq:T_tot} for the parameters
  $\Delta= 3\varkappa$, $\omega_0-\omega_c = -\varkappa$, $g = 0.15 \varkappa$.}
  \label{fig:trans_spec}
\end{figure}

The intensity transmission coefficient for unpolarized light is
  \begin{equation}
    \label{eq:T_tot}
    T=\frac{1}{2}\left(\left|t_{++}^2\right|+\left|t_{+-}^2\right|+\left|t_{-+}^2\right|+\left|t_{--}^2\right|\right).
  \end{equation}
  It does not depend on the electron spin. The same expression describes the transmission of polarized light for unpolarized electron spin. Its frequency dependence is shown in Fig.~\ref{fig:trans_spec}. One can see that in the fast cavity regime $\varkappa \gg g$, the transmission has two Lorentzians peaks at the cavity eigenfrequencies $\omega_c \pm \Delta$ and an asymmetric Fano-like peak at the trion resonance frequency $\omega_0$.

\section{Trion recombination after master equation}
% \section{Master equation solution for the trion recombination}
\label{sec:Lindblad}

Let us describe trion recombination studied in Sec.~\ref{sec:rec} within a different approach similar to Ref.~\onlinecite{Leppenen2022}. The Lindblad equation for the density matrix of the system $\rho$ accounting for the photon escape from the cavity can be written in the form
\begin{equation}\label{eq:Lind}
  \dv{\rho}{t} = -i\qty({\cal H}_{\rm nh}\rho-\rho{\cal H}_{\rm nh}^\dagger) -\sum_{\pm}C_{\pm} \rho C_{\pm}^\dagger,
 \end{equation}
 with the non-Hermitian Hamiltonian and jump operators introduced in Sec.~\ref{sec:model}.
 
In the fast cavity regime, the density matrix elements between trion and electron with photon states are determined by the trion density matrix only:
\begin{subequations}
\begin{multline}
\label{eq:rho_up_p}
	\dot{\rho}_{\Uparrow, \uparrow+} = -[i(\omega_0-\omega_c)+\varkappa] \rho_{\Uparrow, \uparrow +}+\\+i\Delta\rho_{\Uparrow,\uparrow-}+ig\rho_{\Uparrow,\Uparrow},
\end{multline}
\begin{equation}
\label{eq:rho_up_m}
	\dot{\rho}_{\Uparrow, \uparrow-} = -[i(\omega_0-\omega_c)+\varkappa]\rho_{\Uparrow, \uparrow -}+i\Delta\rho_{\Uparrow,\uparrow+}.
\end{equation}
\end{subequations}
Equations for $\rho_{\Uparrow, \downarrow-}$ and $\rho_{\Uparrow, \downarrow+}$ can be obtained by replacing in these equations in the second subscript $\uparrow$, $\Uparrow$, and $\pm$ with $\downarrow$, $\Downarrow$, and $\mp$, respectively. Similarly, the equations for the density matrix elements $\rho_{\Downarrow, \uparrow+}$, $\rho_{\Downarrow, \uparrow-}$, $\rho_{\Downarrow, \downarrow-}$, and $\rho_{\Downarrow, \downarrow+}$ can be obtained from the equations for $\rho_{\Uparrow, \uparrow+}$, $\rho_{\Uparrow, \uparrow-}$, $\rho_{\Uparrow, \downarrow-}$, and $\rho_{\Uparrow, \downarrow+}$ simply by replacing $\Uparrow$ with $\Downarrow$ in the first subscript. The rest of the off-diagonal density matrix elements with the trion states follow from the Hermitian conjugation.

We solve these equations in the steady state and substitute the solution to the equations for the $4\times 4$ density submatrix $\rho_{\rm e-ph}$ in the basis of states $\ket{\uparrow+}$, $\ket{\uparrow-}$, $\ket{\downarrow+}$, $\ket{\downarrow-}$. This gives
 \begin{equation}\label{eq:rhosp}
    \dot{\rho}_{\rm e-ph} = -2\varkappa \rho_{\rm e-ph}-i[{\cal H}_\Delta,\rho_{\rm e-ph}]+\rho_{\rm rec},
  \end{equation}
  where ${\cal H}_\Delta = \Delta (\hat{c}_-^\dagger\hat{c}_++\hat{c}_+^\dagger\hat{c}_-)$ is a part of the Hamiltonian~\eqref{eq:ham} proportional to the cavity modes splitting and the last term $\rho_{\rm rec}$ depends on the trion density matrix
\begin{multline}
	\rho_{\rm rec} = \frac{2\varkappa g^2}{\Delta^2+(\varkappa+\i(\omega_0-\omega_c))^2} \times \\ \times \mqty[{\cal A} \rho_{\Uparrow,\Uparrow}&\frac{\i\Delta}{2\varkappa}\rho_{\Uparrow,\Uparrow} &\frac{\i\Delta}{2\varkappa}\rho_{\Uparrow,\Downarrow} &{\cal A} \rho_{\Uparrow,\Downarrow} \\ -\frac{\i\Delta}{2\varkappa}{\cal B}\rho_{\Uparrow,\Uparrow} & 0 & 0 & -\frac{\i\Delta}{2\varkappa}{\cal B}\rho_{\Uparrow,\Downarrow} \\  -\frac{\i\Delta}{2\varkappa}{\cal B}\rho_{\Downarrow,\Uparrow} & 0 & 0 & -\frac{\i\Delta}{2\varkappa}{\cal B}\rho_{\Downarrow,\Downarrow} \\ {\cal A} \rho_{\Downarrow,\Uparrow}&\frac{\i\Delta}{2\varkappa}\rho_{\Downarrow,\Uparrow} &\frac{\i\Delta}{2\varkappa}\rho_{\Downarrow,\Downarrow} &{\cal A} \rho_{\Downarrow,\Downarrow}],
\end{multline}
with 
\begin{multline}
	{\cal A} = \frac{\Delta^2+(\varkappa+\i(\omega_0-\omega_c))^2}{\Delta^2+\varkappa^2+(\omega_0-\omega_c)^2}{\cal B} = \\ = \frac{\Delta^2+\varkappa^2+(\omega_0-\omega_c)^2}{\Delta^2+(\varkappa-\i(\omega_0-\omega_c))^2}.
\end{multline}
Finally, we solve Eq.~\eqref{eq:rhosp} in the steady state and normalize the result, obtaining
\begin{multline}
	\rho_{\rm e-ph} = \frac{(\omega _0-\omega _c)^2+\varkappa ^2}{(\omega _0-\omega _c)^2+\Delta ^2+\varkappa ^2}\times\\ \times \mqty[\rho_{\Uparrow,\Uparrow} & {\cal G} \rho_{\Uparrow,\Uparrow} & {\cal G} \rho_{\Uparrow,\Downarrow}  & \rho_{\Uparrow,\Downarrow}\\   {\cal G}^* \rho_{\Uparrow,\Uparrow} & \abs{\cal G}^2 \rho_{\Uparrow,\Uparrow}& \abs{\cal G}^2\rho_{\Uparrow,\Downarrow} & {\cal G}^* \rho_{\Uparrow,\Downarrow} \\  {\cal G}^* \rho_{\Downarrow,\Uparrow} & \abs{\cal G}^2 \rho_{\Downarrow,\Uparrow}&\abs{\cal G}^2 \rho_{\Downarrow,\Downarrow} &  {\cal G}^* \rho_{\Downarrow,\Downarrow} \\ \rho_{\Downarrow,\Uparrow} &  {\cal G} \rho_{\Downarrow,\Uparrow} &  {\cal G} \rho_{\Downarrow,\Downarrow}  & \rho_{\Downarrow,\Downarrow}],
\end{multline}
where ${\cal G} = \Delta/(\omega_0-\omega_c-\i \varkappa)$. One can readily check that this is the density matrix of the pure state given in Eq.~\eqref{eq:e-ph}.

\section{Closest orthogonal photon states}
\label{sec:Stokes}

The electron-photon state~\eqref{eq:e-ph} has the same form as for the optically isotropic micropillar cavity except for the replacement of circular photon states $\ket{\pm}$ with $\ket{\tilde{\pm}}$, see Eq.~\eqref{eq:tilde_pm}. These states are generally not orthogonal. The Stockes parameters~\cite{ll2_eng} for them are
\begin{multline}
	 \xi_{1,\pm} = \mp \cos\beta \sin 2\alpha,\quad  \xi_{2,\pm}  = \pm \cos 2\alpha, \\  \xi_{3,\pm} =  \sin \beta \sin 2\alpha.
\end{multline}
The latter parameter is the same for the two states, which evidences their overlap. Generally, by the appropriate rotation of the polarization basis of the Poincaré sphere, one can obtain the Stokes parameters of the form
\begin{equation}\label{eq:stokes1}
    \xi_{1',\pm} = 0,\qquad \xi_{2',\pm} = \pm \xi, \qquad \xi_{3,\pm} = \xi_3
\end{equation}
Where the parameter $\xi = \sqrt{\xi_{1,\pm}^2+\xi_{2,\pm}^2}$ and the polarization basis is rotated about the axis $3$ by the angle $\vartheta$ which is defined by
\begin{equation}
    \cos \vartheta = \frac{\xi_{2,+}}{\sqrt{\xi_{2,+}^2+\xi_{1,+}^2}} = \frac{1}{\sqrt{1+\cos^2 \beta \tan^2 (2\alpha)}}.
  \end{equation}
The closest orthogonal states $\ket{\ttilde{\pm}}$ are defined by the conditions $\bra{\ttilde{\pm}}\ket{\ttilde{\mp}} = 0$ and the overlap $\bra{\ttilde{\pm}}\ket{\tilde{\pm}}$ is maximal. Therefore they should satisfy $\xi_{1',\pm}=0$, $\xi_{2',\pm}=\pm1$, and $\xi_{3',\pm}=0$. Such states are given by Eq.~\eqref{eq:ttildepm}.

\bibliography{clust_cit}

\end{document}